\documentclass[12pt]{article}

\usepackage[margin=1in]{geometry}
\usepackage{import}
\usepackage{format}
\usepackage{algorithm,algorithmic}
\usepackage[backend=biber]{biblatex}

\newcommand{\bS}{\bm{\Sigma}}

\usepackage{xcolor}

\title{\large A Gaussian Process Model of 3D Udder Point Clouds for Teat Length Phenotyping in Dairy Cows}
\author[1]{Maria E. Montes\thanks{montesgonzal@wisc.edu}}
\author[1]{Jo\~ao R. R. Dor\'ea}
\author[2]{Christopher J. Geoga}
\affil[1]{Department of Animal and Dairy Sciences, University of Wisconsin--Madison}
\affil[2]{Department of Statistics, University of Wisconsin--Madison}
\date{}

\begin{document}

\maketitle
\vspace{-3em}

\begin{abstract}
    Phenotyping conformation traits is important for dairy cattle breeding and
    management. Although large-scale phenotyping is possible with 3D imaging
    technologies, manual annotation of anatomical landmarks and long run times
    prevent full pipeline automation. In particular, the morphological
    heterogeneity of cow udders makes automated detection of teat landmarks
    challenging. To address this limitation, we propose and evaluate a method
    for teat length estimation from udder point clouds with a Gaussian process.
    We model the vertical coordinates as the sum of a Gaussian process
    representing the udder floor and an unknown function representing the teat.
    Since the udder floor process is smooth and has a significantly wider
    dependence lengthscale than the teat function, this model allows separating
    the two terms needed for teat landmark definition. To ensure computational
    feasibility, we implement a low-rank approximation of the covariance matrix,
    reducing the computational complexity of the method from $\mathcal{O}(n^3)$
    to $\mathcal{O}(n)$. This approach is both faster and more accurate than
    existing methods, reducing RMSE by a factor of two. It is also more robust
    to uncommon udder morphologies, making it better suited for automated
    phenotyping of large numbers of individuals.
\end{abstract}


\section{Introduction}

High-throughput data and data analytics have driven numerous improvements in
dairy cattle management and genetics \parencite{brito2025}. While farm sensor
systems successfully collect a large volume of routine data, such as milk yield,
breeding records, and health \parencite{brito2025}, collecting conformation
traits remains a challenge on a large scale \parencite{eggerDanner2015}. Udder
conformation traits are heritable and associated with the production performance
and health of dairy cows. Because the teats are the part of the udder in direct
contact with milking equipment, their length and placement can affect the
efficiency of manual \parencite{blake1979} and robotic \parencite{miller1995}
milking cluster attachment. Therefore, teat dimensions are considered for
genetic selection \parencite{miglior2017}, culling decisions
\parencite{Sewalem2004}, and teat cup liner selection \parencite{ronningen1990}.
However, existing teat phenotyping methods, such as caliper
\parencite{batra1984}, ruler \parencite{guarin2016}, or 2D image-based
\parencite{zwertvaegher2011} measurements require physical contact with the
cows, limiting the frequency and scale of teat length records.

Automated measuring methods can ensure consistency in conformation records and
integrate with the farm workflow, thus facilitating large-scale phenotypic data
collection \parencite{fernandes2020}. Recent studies have demonstrated
large-scale phenotyping of body weight \parencite{manzanilla2023} and body
condition score \parencite{mullins2019} using 3D imaging technologies. 3D object
representations allow for estimating size, volume, and area, enabling
contactless body measurements in cows. The process of extracting conformation
traits from 3D point clouds relies on locating key anatomical landmarks. While
annotating these landmarks ensures that derived traits are accurate compared to
live measurements \parencite{fischer2015, lecozler2019, LeCozler2024}, manual
annotation prevents the full automation of the pipeline and its application at
scale.

In recent work, \textcite{montes2026} described a CNN-based pipeline for
processing raw stereo depth images of a ventral view of the udder and obtaining
point cloud representations of each udder quarter. They implemented a
gradient-informed algorithm to isolate the teats and determine their length.
However, this approach relied on rigid cutoffs that do not adapt well to the
significant morphological heterogeneity of cow udders. In this work, we address
this limitation through a method to estimate the length of individual teats from
point clouds using Gaussian processes, which can provide a flexible and accurate
model for automated feature extraction that is applicable to a wide range of
udder morphologies. Critically, since the point clouds obtained from this
pre-processing are very large, we additionally propose computational tools to
reduce the computational burden of teat length estimation from
$\mathcal{O}(n^3)$ to $\mathcal{O}(n)$ work and storage, making this procedure
fast and feasible even on low-power hardware. 

The rest of this work is outlined as follows: Section~\ref{sec:materials and
methods} provides an overview of udder anatomy and describes data acquisition
and pre-processing. We then introduce Gaussian processes and present our
modeling approach for 3D udder point clouds. We present a fast method for
obtaining a low-rank approximation of the covariance matrix, which is critical
for the computational feasibility of the proposed method. This section concludes
with a description of the teat length estimation process and the parameter
optimization pipeline. Finally, Section~\ref{sec:results} compares the estimated
teat lengths with ground-truth lengths and those from a previous method, while
Section~\ref{sec:discussion} discusses the robustness of the proposed model, its
limitations, and future directions.

\section{Materials and Methods}
\label{sec:materials and methods}

\subsection{Udder imaging and pre-processing}
\label{sec:data and data processing}

\begin{figure}[htb]
    \centering
    \begin{subfigure}{.45\linewidth}
        \caption{} \label{fig:udder_anatomy_p1}
        \includegraphics[width=\linewidth]{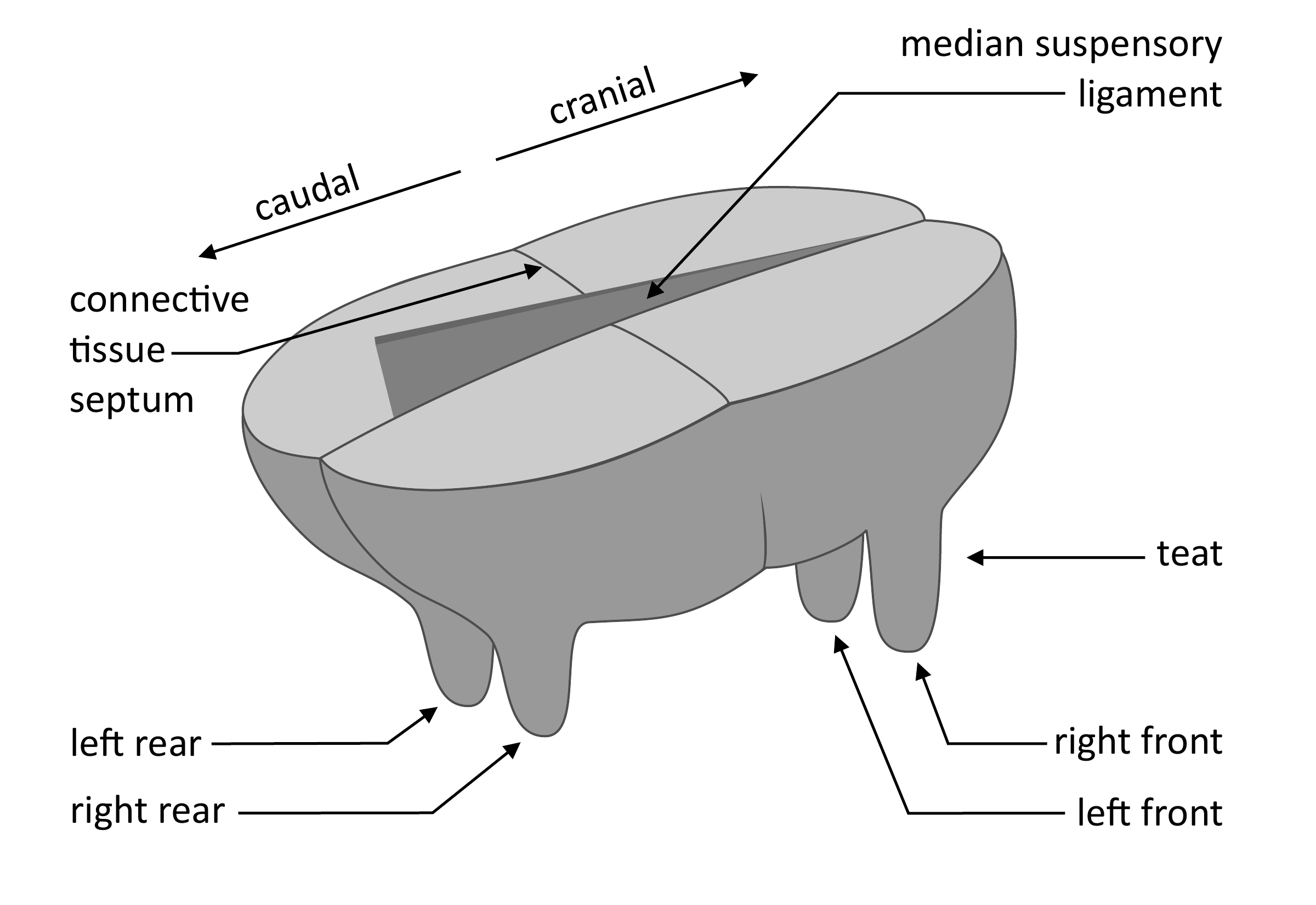}
    \end{subfigure}
    \begin{subfigure}{.45\linewidth}
        \caption{} \label{fig:udder_anatomy_p2}
        \includegraphics[width=\linewidth]{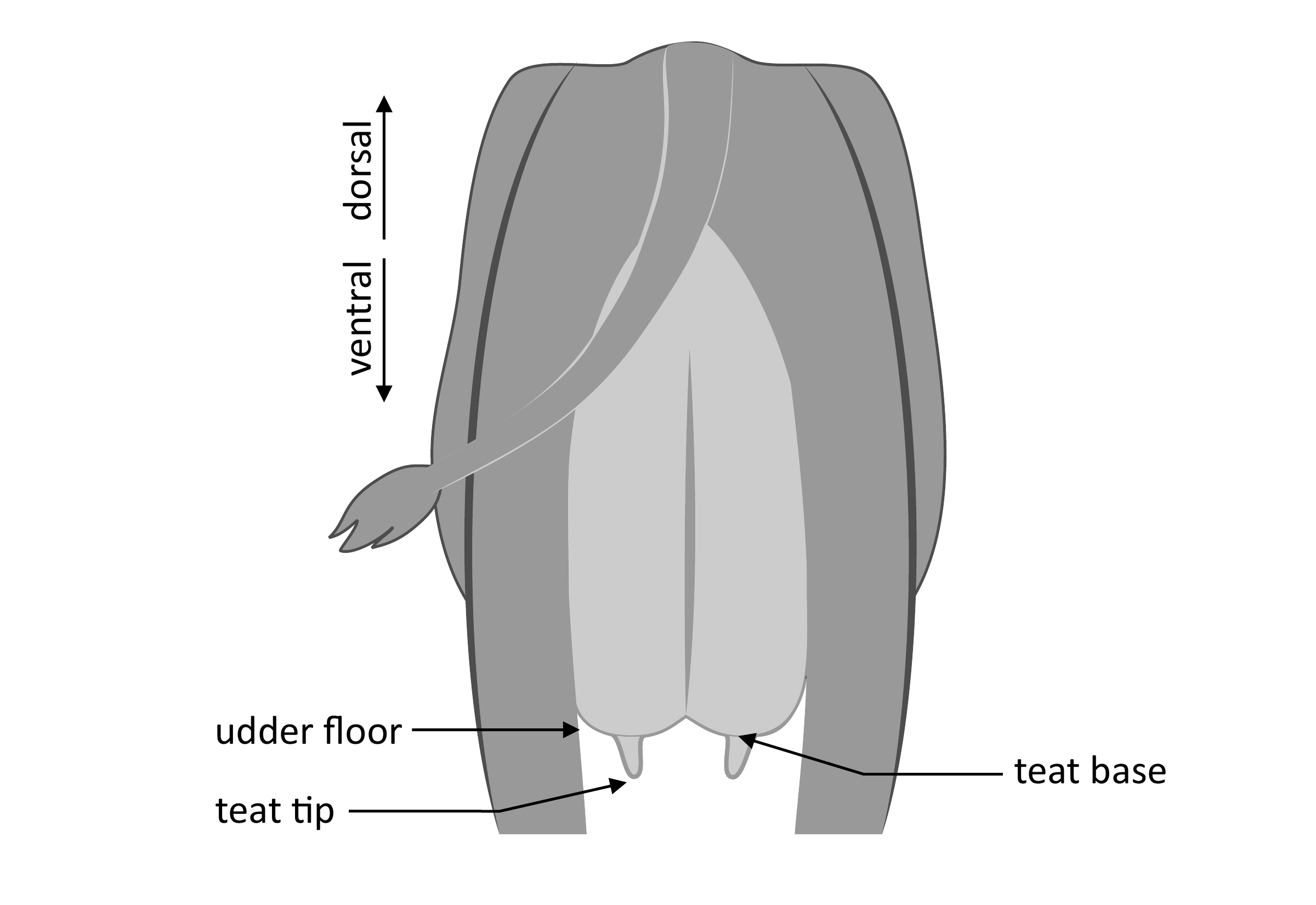}
    \end{subfigure}
    \caption{Cow udder anatomy and morphology. A) Cross-section of the udder
    indicating the major structures. B) Rear view of the udder highlighting the
    teat base, teat tip, and udder floor.}
    \label{fig:udder_anatomy}
\end{figure}

The data used in this work consists of 3D udder point clouds constructed from
depth images. The data processing and modeling choices were driven by the
anatomy and morphology of the udder. A cow’s udder consists of a cluster of four
mammary glands (i.e., quarters) which are supported and attached to the pelvis
by ligaments (Figure~\ref{fig:udder_anatomy_p1}). The median suspensory
ligaments divide the udder into right and left halves, and connective tissues
separate each half into front and rear quarters. Quarters are anatomically
independent and consist of one teat, associated ducts, secretory cells, and
supporting tissues \parencite{nickerson2011}. The teat is a projected structure
that serves as the only exit for the milk. The length of the teat is the
distance from the base of the teat to the tip of the teat
(Figure~\ref{fig:udder_anatomy_p2}). Teats vary in length and placement within
each of the quarters. Front quarters tend to be smaller and have longer teats,
while rear quarters are larger but have shorter teats
\parencite{zwertvaegher2012,weiss2004}. Even though udders are ideally
symmetric, development irregularities and infections result in morphological and
productive differences between right and left halves \parencite{duraes1982}.

\begin{figure}[htb]
    \centering
    \begin{subfigure}{.49\linewidth}
        \begin{subfigure}{.49\linewidth}
            \caption*{A.1 Cow 1061}
            \includegraphics[width=\linewidth]{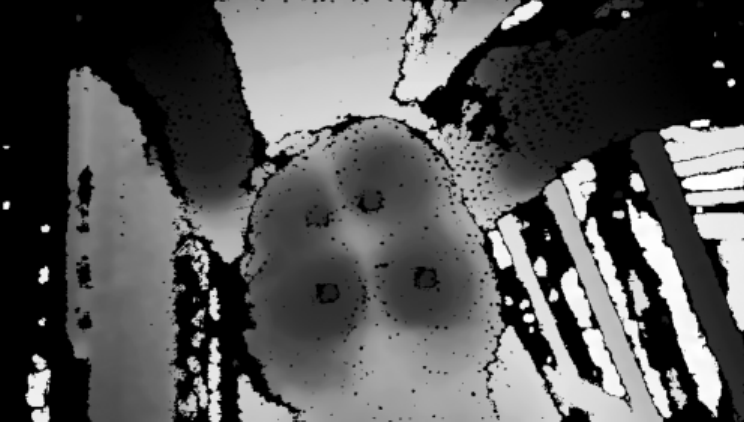}
        \end{subfigure}
        \begin{subfigure}{.49\linewidth}
            \caption*{A.2 Cow 1061}
            \includegraphics[width=\linewidth]{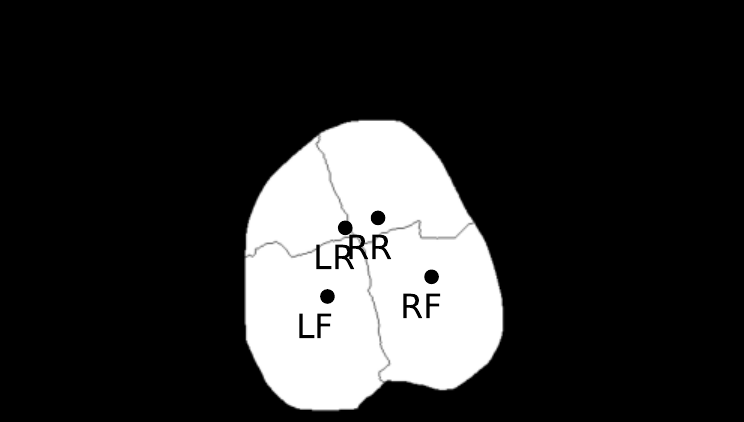}
        \end{subfigure}
        
    \end{subfigure}
    \hfill
    \begin{subfigure}{.49\linewidth}
        \begin{subfigure}{.49\linewidth}
            \caption*{B.1 Cow 1089}
            \includegraphics[width=\linewidth]{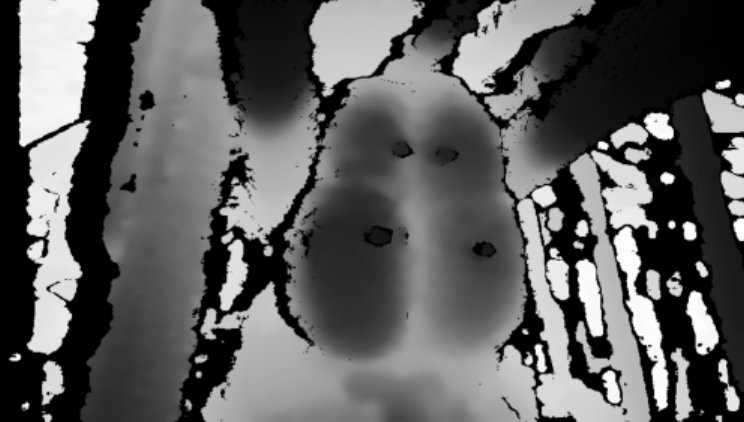}
        \end{subfigure}
        \begin{subfigure}{.49\linewidth}
            \caption*{B.2 Cow 1089}
            \includegraphics[width=\linewidth]{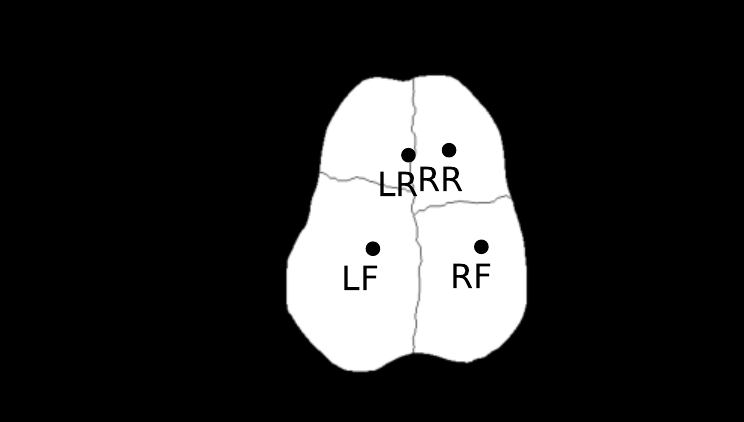}
        \end{subfigure}
    \end{subfigure}
    \begin{subfigure}{.49\linewidth}
        \begin{subfigure}{.49\linewidth}
            \caption*{C.1 Cow 1081}
            \includegraphics[width=\linewidth]{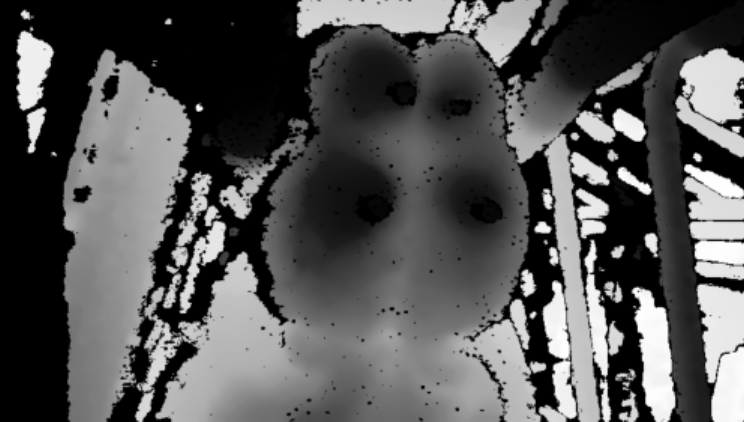}
        \end{subfigure}
        \begin{subfigure}{.49\linewidth}
            \caption*{C.2 Cow 1081}
            \includegraphics[width=\linewidth]{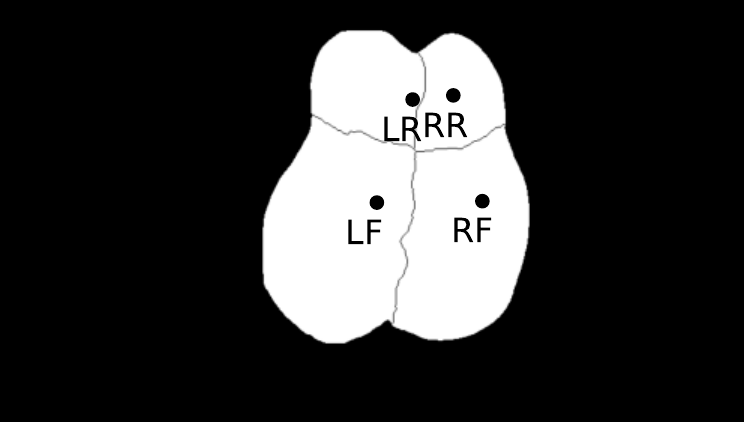}
        \end{subfigure}
    \end{subfigure}
    \hfill
    \begin{subfigure}{.49\linewidth}
            \begin{subfigure}{.49\linewidth}
            \caption*{D.1 Cow 1233}
            \includegraphics[width=\linewidth]{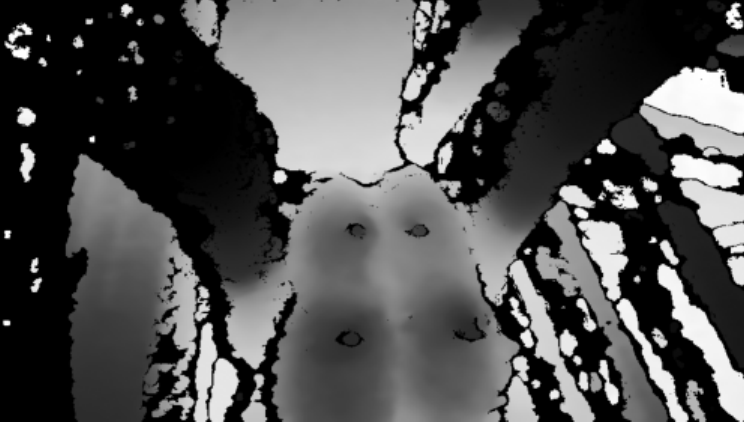}
        \end{subfigure}
        \begin{subfigure}{.49\linewidth}
            \caption*{D.2 Cow 1233}
            \includegraphics[width=\linewidth]{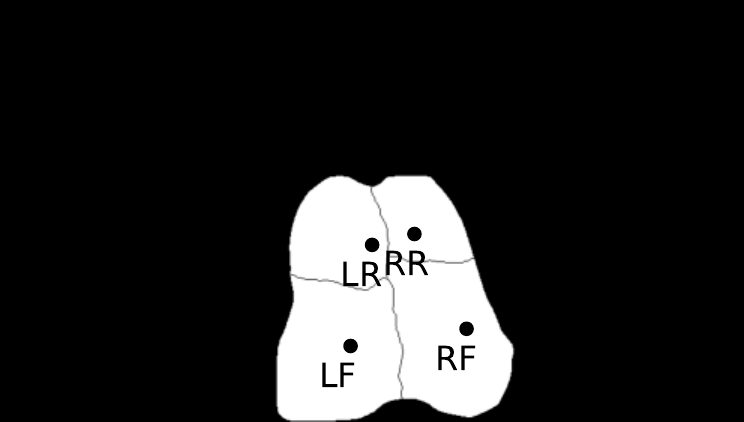}
        \end{subfigure}
    \end{subfigure}
    \caption{Udder depth image processing input (.1) and output (.2) examples (A-D). The input is a depth image and the output is a segment of the udder contour and the teat locations labeled as right front (RF), left front (LF), right rear (RR), and  left rear (LR).}
    \label{fig:input_output}
\end{figure}

In image analysis, segmentation is the process of partitioning the image into
independent regions. The pipeline from  \textcite{montes2026}, which integrates
convolutional neural networks (CNN) and image processing tools, takes a raw
depth image of the udder as input and returns the quarter segmentation of the
udder (Figure \ref{fig:input_output}). \textcite{montes2026} collected depth
videos from the udders of 150 lactating Holstein cows (parity 2.2 ± 1.2, days in
milk: 155 ± 103) at a commercial dairy farm with an Automated Milking System
(DeLaval VMS 3.0, Tumba, Sweden). They used a RealSense 455 depth camera at 30
frames per second with a resolution of 848 x 480 pixels to record a ventral view
of the udder before milking. 

\begin{figure}[htb]
    \centering
    \includegraphics[width=\linewidth]{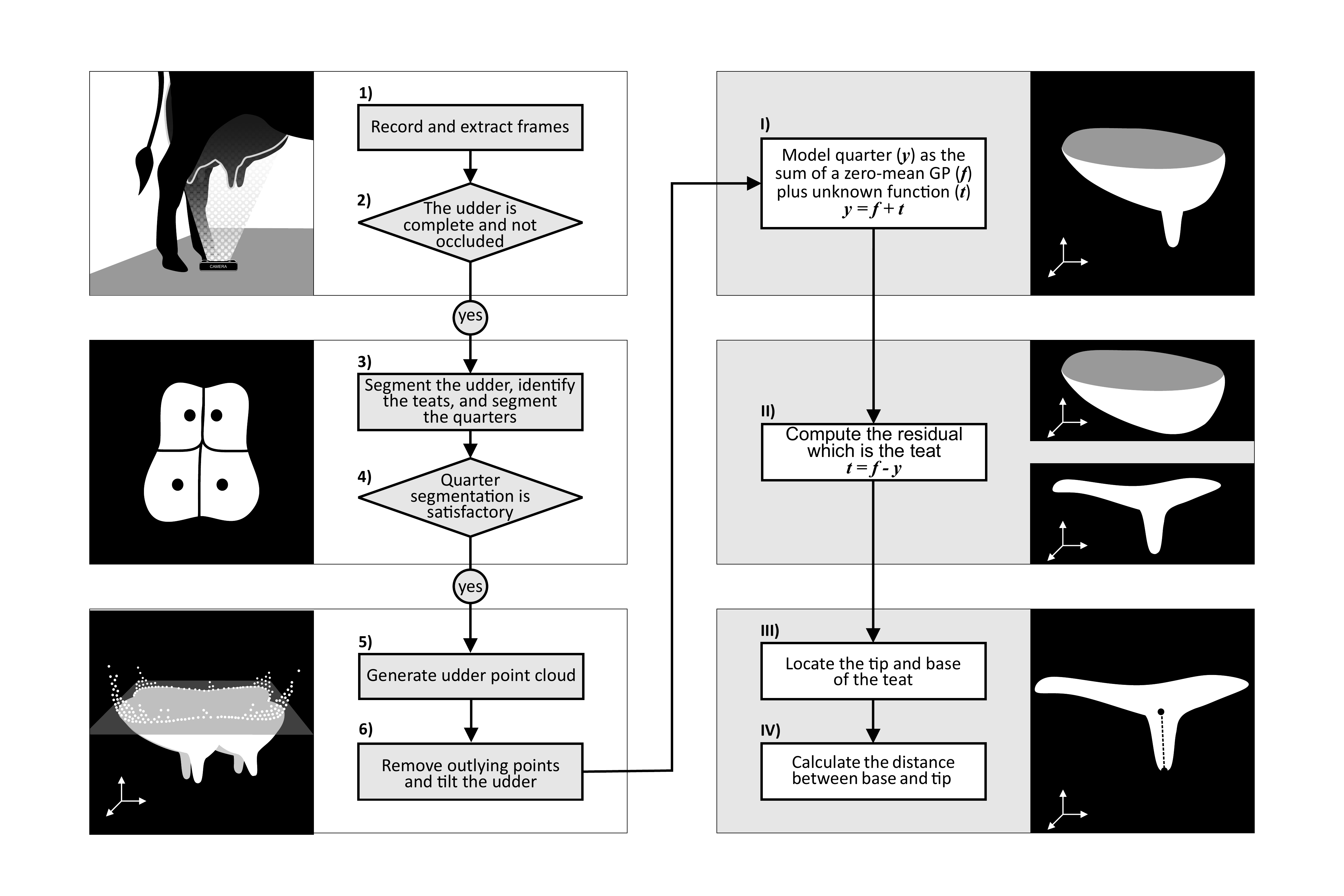}
    \caption{
    Udder depth image processing: steps 1-6 implement CNN models and image
    processing tools to obtain the quarter segmentation from an udder depth
    image, and generate and clean the udder point cloud from the depth image.
    Teat length estimation: steps I-IV use a Gaussian process to model the udder
    quarter, separate the udder floor from the teat, and estimate teat length
    from the point cloud.}
    \label{fig:flowchart}
\end{figure}

In this work, we selected twelve cows with varying levels of udder symmetry from
the dataset of \textcite{montes2026} to extract teat lengths from udder point
clouds. We obtained point cloud representations of udder quarters using the
procedure described by \textcite{montes2026} (Figure \ref{fig:flowchart}). The
pipeline first uses a CNN classifier to select images where the udder is not
occluded. It then identifies the udder region using an instance segmentation CNN
and locates the four teats using a keypoint detection CNN. The Watershed
transform \parencite{gonzalez2017} is then used to segment the udder into
quarters using teat locations as seeds, and poor segmentation results are
discarded using a CNN classifier.  The depth images with good segmentation are
then converted into point clouds using camera intrinsics. Finally, outlying
points are removed, and the point cloud is rotated so that the plane defined by
the tips of the three lowest teats is horizontal.

\subsection{Gaussian Processes}
\label{sec:GP and teatlength}

To automate teat length measurements, it is necessary to define a model of the
udder surface, and in this work we use a Gaussian process. A Gaussian process
$Z(\bm{x}) \; ; \; \bm{x} \in \Omega\subset \mathbb{R}^d$ is a stochastic process
for which all finite-dimensional marginals have multivariate Gaussian
distributions. Since the multivariate normal distribution is fully specified by
its first two moments, the process $Z(\bm{x})$ is specified by its \emph{mean}
and \emph{covariance} functions, denoted by
\begin{align} 
    \mu(\bm{x}) &= \mathbb{E}Z(\bm{x}) \label{mean_fun} \\
    K(\bm{x},\bm{x}') 
    &= 
    \mathbb{E}[(Z(\bm{x}) -\mathbb{E}Z(\bm{x}))(Z(\bm{x'}) -\mathbb{E}Z(\bm{x'}))] 
    \label{cov_fun},
 \end{align}
as this implies that any finite set of observations $y_i = Z(\bm{x}_i)$ at
locations $\{\bm{x}_i  : \bm{x}_i \in \Omega\}_{i=1}^n$ has a multivariate
normal distribution $\bm{y} \sim \mathcal{N}(\bm{\mu}, \bm{\Sigma})$ with $\mu_j
= \mu(\bm{x}_j)$ and $\bm{\Sigma}_{j,k} = K(\bm{x}_j, \bm{x}_k)$. The covariance
function $K(\bm{x}, \bm{x}')$ must be positive definite, meaning that the
corresponding matrix $\bm{\Sigma}$ is positive (semi-)definite for any
collection of measurement locations $\{\bm{x}_j\}_{j=1}^n$. Standard choices for
$K(\bm{x}, \bm{x}')$ are the \emph{squared exponential} model $K(\bm{x},
\bm{x}') = \sigma^2 \exp(-\rho^{-1} || \bm{x} - \bm{x}' ||^2)$, or the Mat\'ern
model (of which the squared exponential is a special case) given by $K(\bm{x},
\bm{x}') = \sigma^2 \mathcal{M}_{\nu}(\rho^{-1} || \bm{x} - \bm{x}' ||)$, with
$\mathcal{M}_{\nu}$ being the Mat\'ern correlation function given by
\begin{align}
    \mathcal{M}_{\nu}(r) =
    &\frac{2^{1-\nu}}{\Gamma(\nu)}\left(\frac{\sqrt{2\nu}r}{\rho}\right)^\nu
    K_\nu\left(\frac{\sqrt{2\nu}r}{\rho}\right), \label{matern}
\end{align}
where $\Gamma$ is the gamma function $K_\nu$ is the modified second-kind Bessel
function \parencite{nist}.  One typically refers to parameters $\sigma^2$,
$\rho$, and $\nu$ and the \emph{scale}, \emph{range}, and \emph{smoothness}
parameters respectively. While the interpretation of $\sigma^2$ and $\rho$ is
natural, controlling the marginal variance and correlation lengthscales
respectively, the parameter $\nu$ controls the \emph{mean-square
differentiability} \parencite{stein1999}, an important characteristic of the
process $Z(\bm{x})$ that has significant implications on, e.g., the spectral
decay properties of $\bm{\Sigma}$.

\subsubsection{Udder model}
The main goal of this work is to obtain teat length from the point cloud of the
udder's ventral surface. Obtaining teat length requires locating the tip and the
base of the teat. Although locating the teat tip can be conceptually
straightforward because it usually corresponds to the lowest point within the
quarter, locating the base, where the teat and the udder floor meet, represents
a bigger challenge. Defining the teat base requires identifying both structures.
To that aim, we model the vertical position of the 3D points as a function of
location, where the vertical position is the sum of the udder floor and the
teat. Since quarters from the same udder can vary in shape and size and can be
separated by the CNN pipeline described above, we model each quarter
independently. 

We model vertical position $y(\bm{x})$ for locations $\{\bm{x}_i\mid \bm{x}_i
\in \mathbb{R}^2\}_{i=1}^n$ as the sum of a zero mean Gaussian process
$f(\bm{x})$ and an unknown function representing the teat, which we denote with
$t(\bm{x})$:
\begin{align}
    y(\bm{x}) =& \underbrace{f(\bm{x})}_{\text{udder floor}} + \; \; \underbrace{t(\bm{x})}_{\text{teat}} \label{udder_eq} \\ 
    f(\bm{x}) \sim \;& \mathcal{GP}(0, K(\bm{x}, \bm{x}')). \nonumber 
\end{align}
The key design choice that makes this model useful is that the udder floor
process $f(\bm{x})$ is smooth and has a significantly wider dependence
lengthscale than the teat function $t(\bm{x})$, making it possible to separate
the two terms' contributions to the measured data $\bm{y}$.

In particular, for a given udder point cloud, let $\bm{y}=
[y(\bm{x_1}),y(\bm{x_2}), \dots, y(\bm{x_n})]^T$ be the observed vertical
positions of the points located within one quarter and $\bS_{j,k} = K(\bm{x}_j,
\bm{x}_k)$ be the covariance matrix of $\bm{f} = [f(\bm{x}_1), ...,
f(\bm{x}_n)]$. Letting $\bS = \bm{Q} \bm{\Lambda} \bm{Q}^T$ be the
eigendecomposition of $\bS$ sorted so that $\Lambda_{j,j} \geq \Lambda_{k,k}$
for $k > j$, we note that if $\Lambda_{r, r}$ is small (say, below machine
epsilon of $\approx 10^{-16}$), one may represent $\bm{f}$ to machine precision
with the truncated representation $\bm{f} = \bm{Q}_{r} \bm{s}$, where $\bm{s}
\sim \mathcal{N}(0,\bm{\Lambda}_{r})$ and the subscripts $\bm{Q}_r$ and
$\bm{\Lambda}_r$ denote keeping the first $r$ columns and the principal $r
\times r$ submatrix respectively.

If $\bm{t} = [t(\bm{x}_1), ..., t(\bm{x}_n)]$, the contribution to $\bm{y}$ from
the unknown teat function, is nearly orthogonal to the columns of $\bm{Q}_r$,
then one has that
\begin{equation*} 
  \hat{\bm{t}} 
  =
  (\mathcal{I} - \bm{Q}_r \bm{Q}_r^T) \bm{y} 
  = 
  (\mathcal{I} - \bm{Q}_r \bm{Q}_r^T) (\bm{f} + \bm{t}) 
  =
  (\mathcal{I} - \bm{Q}_r \bm{Q}_r^T) (\bm{Q}_r \bm{s} + \bm{t}) \approx  \bm{t}.
\end{equation*}
Fortunately, the fact that $f(\bm{x})$ varies on such wider scales as
$t(\bm{x})$ makes such a separating decomposition easy: if one takes the model
that $K(\bm{x}, \bm{x}')$ corresponds to a very smooth process with a large
range parameter, then the sharply peaked teat will not live in the
non-degenerate column space of $\bS$.

Moreover, as we will demonstrate next, choosing such a smooth and strongly
dependent process also means that this estimator can be computed extremely
rapidly, even with data sizes $n \approx 15,000$ that would otherwise pose
numerical issues both in terms of runtime and storage cost as well as in terms
of numerics.

\subsubsection{Fast methods for obtaining $\bm{Q}_r$}

A na\"ive computation of $\bm{Q}_r$ requires a full eigendecomposition of the
covariance matrix $\bm{\Sigma}$, which scales as $\mathcal{O}(n^3)$ and has a
much higher prefactor than, for example, a Cholesky factorization, and repeated
computation of an eigendecomposition for each quarter of the udder with $n
\approx 15,000$ across potentially hundreds of different cows is obviously too
computationally expensive to be feasible. While there are several potential
approaches to avoid ever completely factorizing---or even
\emph{assembling}---the matrix $\bm{\Sigma}$, in this work we will employ the
simplest approach that can be implemented in tens of lines of code.

\begin{figure}[!ht]
  \centering
  \includegraphics[width=\textwidth]{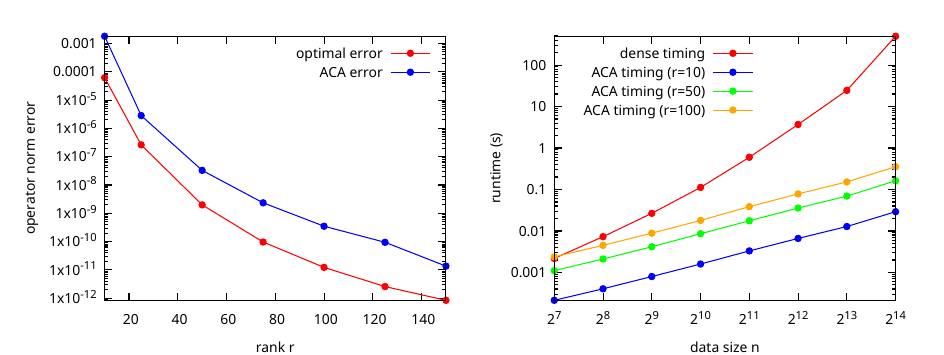}
  \caption{An error (left panel) and runtime (right panel) comparison of the
  non-oversampled adaptive cross approximation (ACA) as a method for extracting
  rank-$r$ approximations of a Mat\'ern covariance matrix using the covariance model
  employed in this work, demonstrating the ability of the ACA to extract the $r
  \approx 150$-dimensional non-degenerate column space basis $\bm{Q}_r$ in under
  one second for the data sizes used in this work.}
  \label{fig:aca}
\end{figure}

The \emph{adaptive cross approximation} (ACA) with partial pivoting
\parencite{bebendorf2003} is an algorithm for assembling low-rank approximations
of kernel matrices for which the kernel---at least in the region in which is
being evaluated---is smooth. It was originally used in the hierarchical matrix
literature \parencite{hackbusch2015} to compress off-diagonal blocks of kernel
matrices that arise in the numerical solution of partial differential equations,
but its applicability has extended far beyond that
\parencite{townsend2013,arora2026}.
Functionally, it mimics a partially pivoted LU decomposition
\parencite{trefethen2022}, greedily picking pivots based only on the current row
and column information, and by doing so reduces the cost of assembly rank $k$
approximation of an $n \times n$ matrix from $\mathcal{O}(n^3)$ work to
$\mathcal{O}(k n)$. While it is sub-optimal in the sense of not providing
rank-$r$ approximations whose error is exactly the $r+1$-th singular value, it
does come with theoretical guarantees of convergence in some settings with
sufficiently smooth kernel functions and favorable geometry
\parencite{hackbusch2015}. As Figure \ref{fig:aca} demonstrates, the cost of
using an ACA for low-rank approximation for smooth covariance functions in terms
of efficiency is small, but the gains in performance are significant.

Due to this favorable runtime cost and accuracy tradeoff, we use the ACA is used
to assemble the approximation $\bm{\Sigma} = \bm{U} \bm{V}^T$, where $\bm{U}$
and $\bm{V}$ are matrices of size $n \times r(\epsilon)$ with $r(\epsilon)$ the
numerical rank of $\bm{\Sigma}$ at an error tolerance of $\epsilon = 10^{-12}$.
Obtaining a basis for the non-degenerate column space $\bm{Q}_r$ from such a
$\bm{U}$ and $\bm{V}$ can be done by simply converting $\bm{U} \bm{V}^T$ into a
partial QR factorization by computing $\bm{U} = \bm{Q}_1 \bm{R}_1$, $\bm{R}_1
\bm{V}^T = \bm{Q}_2 \bm{R}_2$, and observing that $\bm{U} \bm{V}^T = \bm{Q}_1
\bm{Q}_2 \bm{R}_2$, and so $\bm{Q}_r = \bm{Q}_1 \bm{Q}_2$.  Since all of the
intermediate QR factorizations are done on matrices of dimension $n \times r$
and obtaining $\bm{\Sigma} = \bm{U} \bm{V}^T$ costs $\mathcal{O}(n)$ work in
this regime, the total runtime cost of obtaining $\bm{Q}_r$ has been reduced
from $\mathcal{O}(n^3)$ work to $\mathcal{O}(n)$. A summary of the ACA algorithm
for reference is given in Appendix \ref{app:aca}.

\subsection{Trigonometric post-processing for teat length}
\begin{figure}[htbp]
    \centering
    \begin{subfigure}{.45\linewidth}
        \caption{} \label{fig:teat_len1}
        \includegraphics[width=\linewidth]{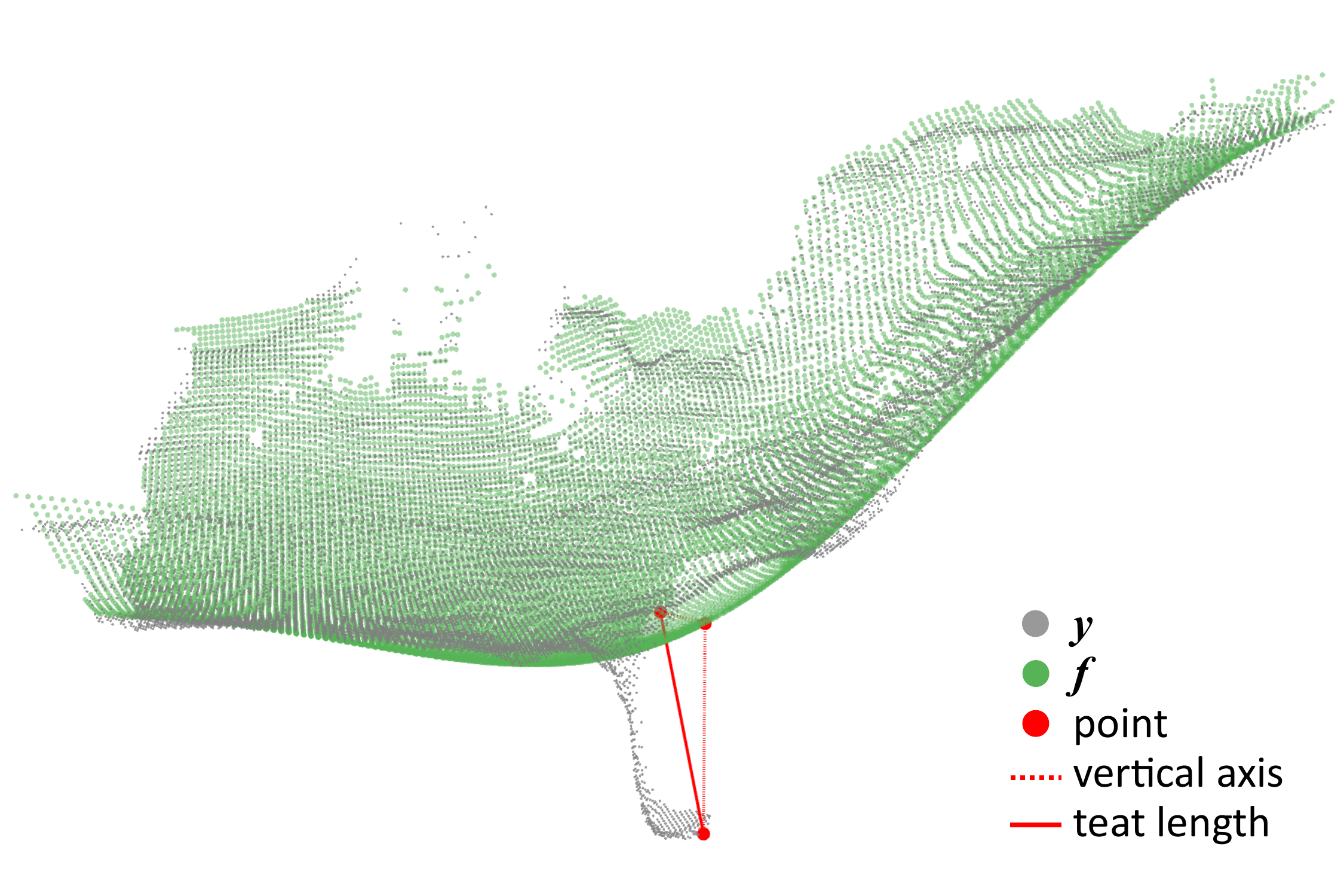}
    \end{subfigure}
    \begin{subfigure}{.445\linewidth}
        \caption{} \label{fig:teat_len2}
        \includegraphics[width=\linewidth]{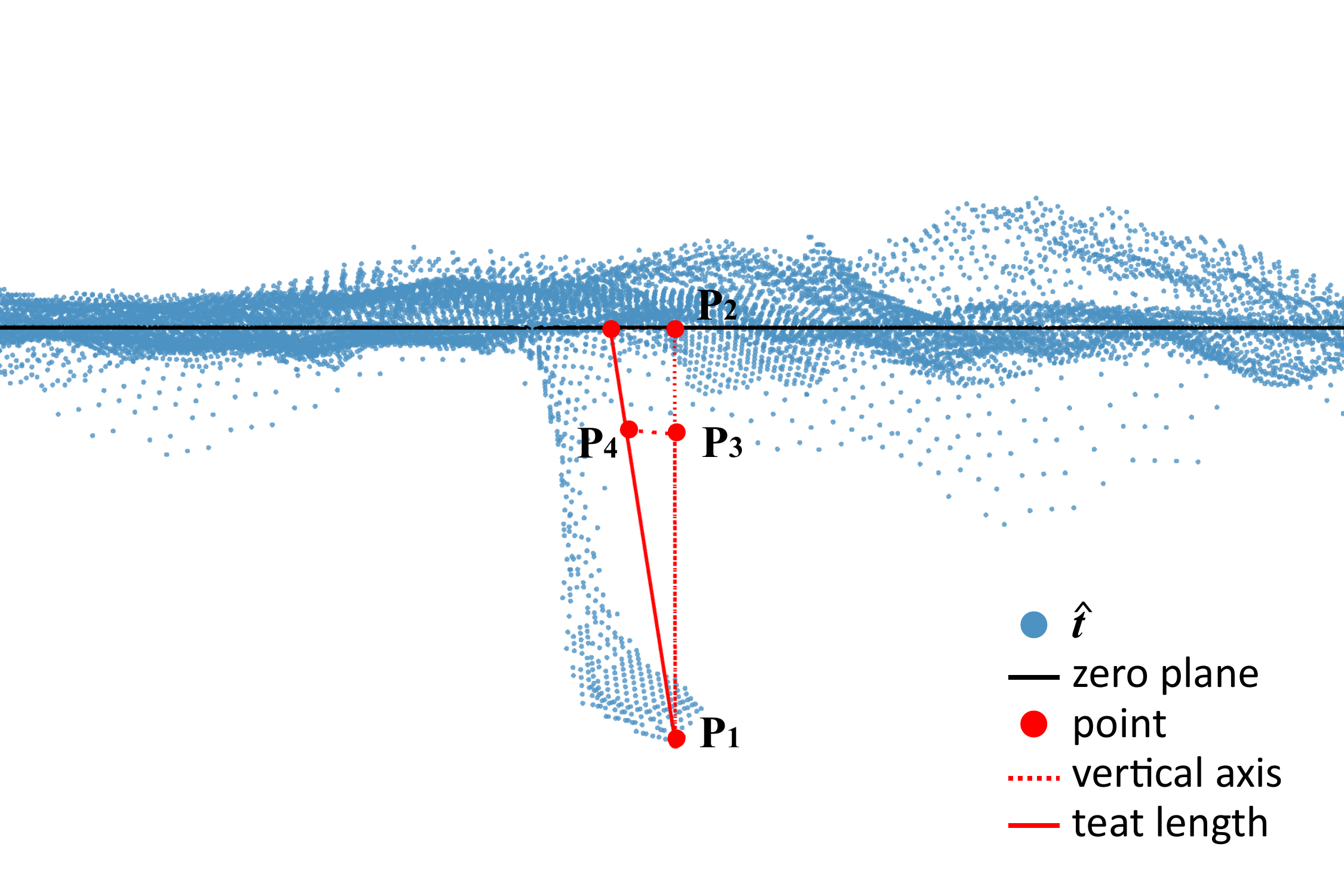}
    \end{subfigure}
    \caption{
     Point cloud of an udder quarter indicating the elements of the udder model
     and teat length estimation. A) Observed udder surface $\bm{y}$ and the
     corresponding smooth Gaussian process $\bm{f}$. B) Teat length estimation
     $\hat{d}$ derived from $\hat{\bm{t}}$. Teat length is the hypotenuse along
     the teat axis $(\mathbf{P}_4 - \mathbf{P}_1)$ corresponding to the vertical
     segment $(\mathbf{P}_2 - \mathbf{P}_1)$. The landmarks include the teat tip
     $\mathbf{P}_1$, its vertical projection onto the zero plane $\mathbf{P}_2$,
     and $\mathbf{P}_4$ which is used to define the teat orientation.
    }
    \label{fig:teat_len}
\end{figure}

Once we have obtained $\hat{\bm{t}}$, we employ a heuristic method for
estimating teat length, $\hat{d}$, based on basic trigonometric arguments
(Algorithm~\ref{alg:teat_len}). Conceptually, the approach is based on the
observation that teats are often angled relative to the udder floor.
Mathematically, we make this approach concrete in the following procedure based
on several geometric reference points (Figure \ref{fig:teat_len}). The first
point, $\mathbf{P}_1$, represents the teat tip and is located at the horizontal
coordinate $\bm{x}^*$ where $\hat{t}$ has its minimum value. The second point,
$\mathbf{P}_2$, has the same horizontal coordinate $\bm{x}^*$ but has vertical
position zero. The third point, $\mathbf{P}_3$, is also located at $\bm{x}^*$
with a vertical position set to one-fourth of the tip's height. The teat radius,
$R_\text{teat}$, is the minimum distance between $\mathbf{P}_3$ and surface
points in $\hat{\bm{t}}$. The fourth point, $\mathbf{P}_4$, is a point on the
teat length axis located at the median horizontal position of the points in
$\mathcal{X}$ that are vertically below $\mathbf{P}_3$ and within
$R_{\text{teat}}$ with vertical position set to one-fourth of the tip's height.
Finally, we calculate teat length $\hat{d}$ as the hypotenuse along the teat
axis $(\mathbf{P}_4 - \mathbf{P}_1)$ corresponding to the vertical segment
$(\mathbf{P}_2 -\mathbf{P}_1)$, thereby accounting for the teat inclination
relative to the udder floor.

\begin{algorithm}
    \caption{Teat length estimation from 3D point clouds}
    \label{alg:teat_len}
    \begin{algorithmic}[1]
    \REQUIRE Parameters $\theta$, and $\bm{y}$ defined over spatial coordinates $\bm{x}$ 

        \STATE $\bm{Q}_r, \bm{R} \approx \bm{\Sigma}_{\bm{\theta}}$
        \COMMENT{ACA-accelerated factorization}

        \STATE $\hat{\bm{t}}_{jk}(\theta)  \gets (\mathcal{I} - \bm{Q}_{r}(\theta) \bm{Q}_{r}(\theta)^T) \bm{y}$ \COMMENT{Segment the teat}
        
        \STATE $\bm{x}^* \gets \argmin_{\bm{x}} \hat{t}(\bm{x}) $ \COMMENT{Locate the tip}
        
        \STATE 
        $\mathbf{P}_1 \gets \begin{bmatrix} \bm{x}^* & \hat{t}(\bm{x}^*) \end{bmatrix}^T$
        \STATE $\mathbf{P}_2 \gets \begin{bmatrix} \bm{x}^* & 0 \end{bmatrix}^T$
        \STATE $\mathbf{P}_3 \gets \begin{bmatrix} \bm{x}^* & \frac{1}{4}\hat{t}(\bm{x}^*) \end{bmatrix} ^T$
        
        \STATE $R_{\text{teat}} \gets \min_{\bm{x}} \left\lVert \begin{bmatrix} \bm{x} & \hat{t}(\bm{x}) \end{bmatrix}^T - \mathbf{P}_3 \right\rVert$ \COMMENT{Define teat radius}
        
        \STATE $\mathcal{X} \gets \left\{ \bm{x} \mid \lVert \bm{x} - \bm{x}^* \rVert < R_{\text{teat}} \;\text{and}\; \hat{t}(\bm{x}) < \tfrac{1}{4} \hat{t}(\bm{x}^*) \right\}$ \COMMENT{Subset points}
        
        \STATE $\mathbf{P}_4 \gets \begin{bmatrix} \text{median}_{\bm{x} \in \mathcal{X}}(\bm{x}) &  \tfrac{1}{4} \hat{t}(\bm{x}^*) \end{bmatrix}^T$ 
        
        \STATE $\hat{d} 
        \gets \frac{\lVert \mathbf{P}_2-\mathbf{P}_1 \rVert^2 \lVert \mathbf{P}_4-\mathbf{P}_1 \rVert}{(\mathbf{P}_2-\mathbf{P}_1) \cdot (\mathbf{P}_4-\mathbf{P}_1)}$ \COMMENT{Calculate length}

        \RETURN $\hat{d}$
    \end{algorithmic}
\end{algorithm}

\subsection{Parameter optimization}

In order to select parameters for the Mat\'ern kernel used to identify the udder
function, we define a loss function based on the discrepancy between this
predictive approach described above and labeled udder scans. In particular, for
teat length $d_{jk}$, a lactation physiology expert blinded to the algorithm's
outputs manually annotated the teat tip and base locations on the point clouds
from 12 cows (48 quarters in total). We randomly split these quarters into a
testing set of 4 cows and a training set of $n=8$ cows, grouping by cow. This
labeled data gives a natural loss function of

\begin{align}
    \mathcal{L}(\theta) = \sum_{j =1}^n\sum_{k =1}^4 \left(d_{jk} -
    \hat{d}_{jk}(\theta)\right)^2,
\end{align}

where $\theta = (\rho, \nu)$ are the range and smoothness parameters of the
Mat\'ern covariance function, and $\hat{d}_{jk}(\theta)$ is obtained from
Algorithm~\ref{alg:teat_len}. We fixed the scale parameter to one while treating
the range and smoothness parameters as variables. We then minimized this loss
function using the nonlinear optimization library NLopt \parencite{NLopt} using
automatic differentiation to obtain gradients \parencite{GMSS_2022}.


\section{Results} 
\label{sec:results}

\begin{figure}[htb]
    \centering
    \includegraphics[width=0.85\linewidth]{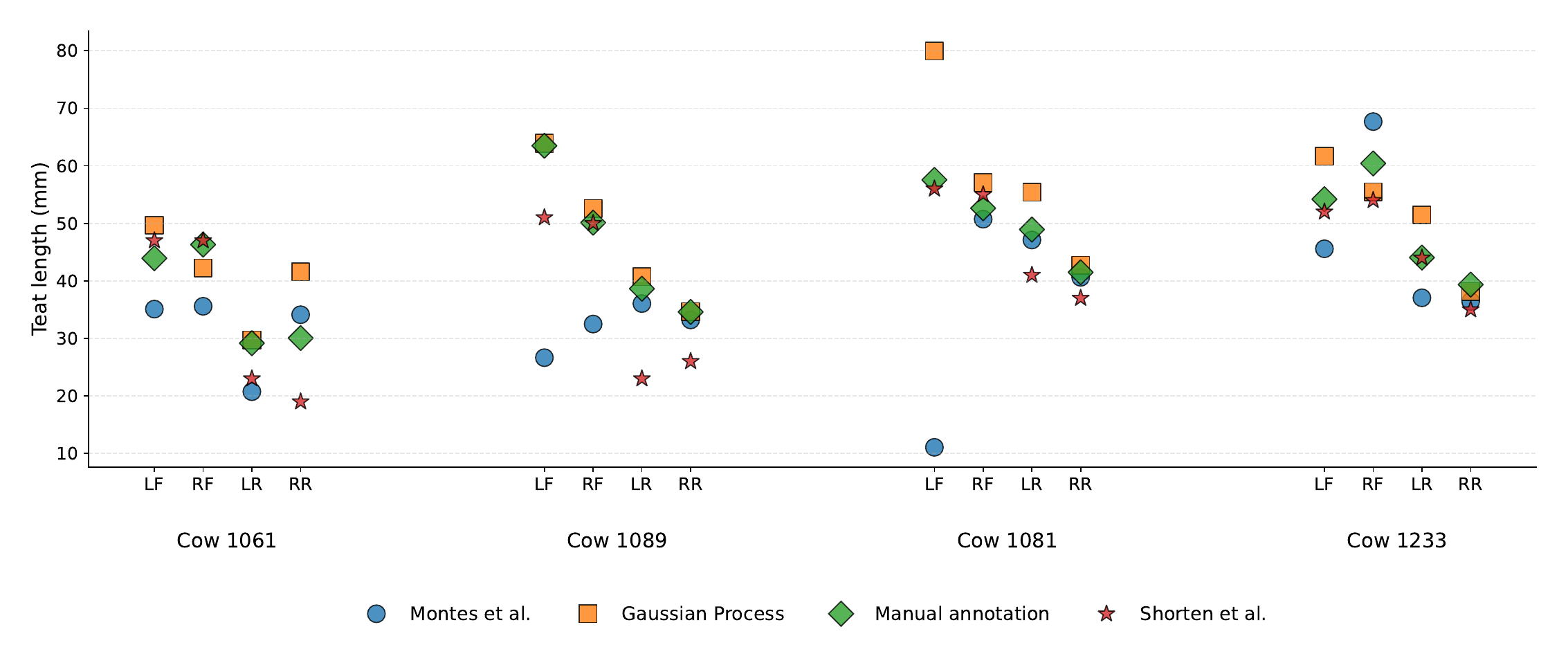}
    \caption{Comparison of teat length estimates on test quarters for the three
    measurement methods: Montes et al. (2026, blue circles), Shorten et al.
    (2021, red stars), the Gaussian Process method (orange squares), and Manual
    annotation (green diamonds). Quarters are labeled as right front (RF), left
    front (LF), right rear (RR), and  left rear (LR).} 
    \label{fig:error_plots}
\end{figure}

The Gaussian process approach provided more accurate teat length estimates than
the methods described by \textcite{montes2026} and our implementation, detailed
in Appendix~\ref{app:shorten}, of the method described by \textcite{shorten2021}
when compared to manual annotations, in a fraction of the computational time
(Table~\ref{tab:rmse}). The Root Mean Square Error (RMSE) values for the test
and training quarters of the Gaussian process were consistent and outperformed
those of the other two methods. Figure~\ref{fig:error_plots} displays the teat
length estimates on the test quarters compared to manual annotation alongside
both the \textcite{montes2026} and \textcite{shorten2021} methods, demonstrating
that most of the teat length estimates from the Gaussian process method align
with manual annotations.

\begin{table}[htb]
    \centering
    \caption{Performance for automated teat length estimation compared to manual annotations.}
    \begin{tabular}{l c c c c}
        \hline
        Method & set* & Quarters ($n$) & RMSE (mm) & Runtime / quarter (s)\\
        \hline
        \textcite{montes2026} & full & 48 & 16.32 & 312\\
        \textcite{shorten2021} & full & 48 & 14.77 & 172\\
        Gaussian process & train & 32 & 7.69 & 12\\
        & test & 16 & 7.48 & -\\
        \hline
        \multicolumn{5}{l}{\small *Sets used for parameter optimization in the Gaussian process method}
    \end{tabular}
    
    \label{tab:rmse}
\end{table}

\begin{figure}[htb]
    \centering
    \begin{subfigure}{0.9\linewidth}
        \caption{Cow 1081} \label{fig:1081}
        \includegraphics[width=\linewidth]{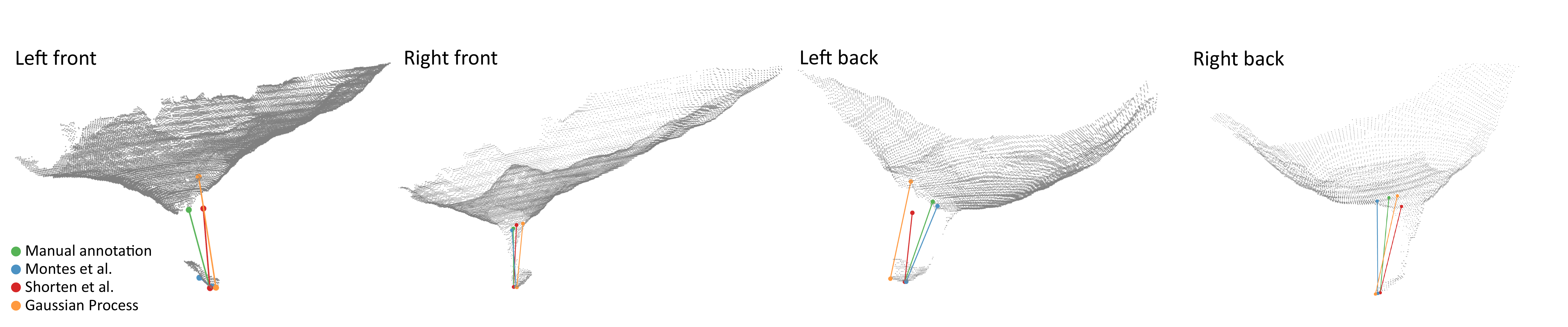}
    \end{subfigure}
    \begin{subfigure}{0.9\linewidth}
        \caption{Cow 1089} \label{fig:1089}
        \includegraphics[width=\linewidth]{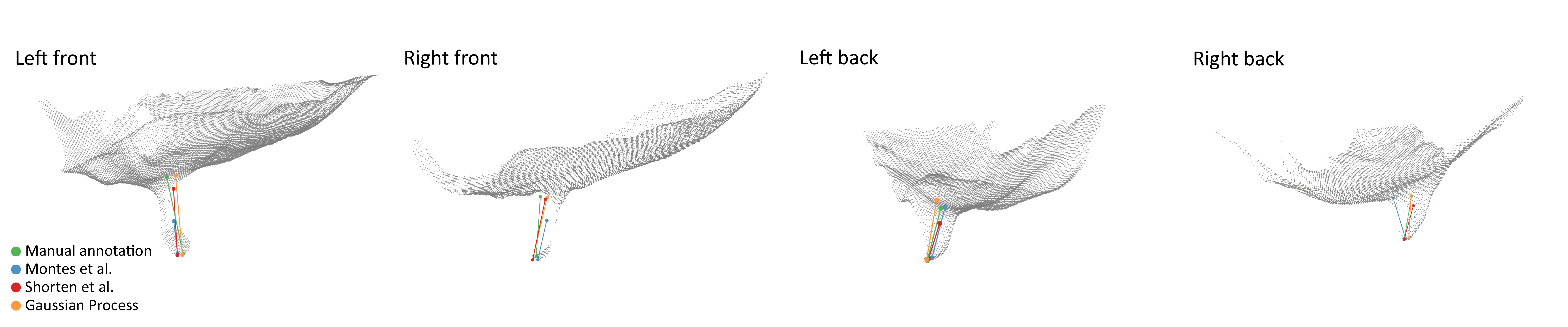}
    \end{subfigure}
    \caption{Point cloud representations of four udder quarters with estimated
    teat length vectors for manual annotation (green), Montes et al. (2026;
    blue), Shorten et al. (2021; red), and the Gaussian process method
    (orange).}
    \label{fig:teat_len_examples}
\end{figure}

The improved performance of the Gaussian process method was evident in
challenging quarters with angled teats or missing data in the point cloud. The
alternative methods and the proposed method perform similarly for straight teats
and clean scans. However, the \textcite{montes2026} approach consistently
underestimates teat length in non-ideal quarters. Similarly, the method of
\textcite{shorten2021} mostly provides accurate lengths, but yields inaccurate
lengths that substantially deviate from manual annotation when certain features
of the teat are not captured in the 3D point scans, for example in the left
front quarter (LF) of cow 1081. The Gaussian process approach was the only one
of the three to provide an accurate length estimate. Similarly, the Gaussian
process tip and base locations remained robust on the left front (LF) teat of
cow 1089, despite the teat being angled with respect to the udder floor (Figure
~\ref{fig:1089}). In contrast, there was no major difference in teat length
between manual annotation and the automated methods on quarters where the point
clouds were mostly complete, and teats were close to perpendicular to the udder
floor, such as the right front (RF) and right rear (RR) quarters of cow 1081.

The key to this robustness is that the function being interpolated is the udder
function, which is presumed to be smooth. Since it is being modeled as a GP with
a Mat\'ern covariance function with a large range parameter, the points observed
across the udder surface are sufficient to characterize the udder floor below
the teat and make interpolation very stable. The teat function, while smooth in
three dimensions, may have two-dimensional projections with very sharp behavior
depending on the angle of the camera and variations in udder morphology, and so
this method approaches teat length estimation after removing the underlying
udder function from a purely geometric perspective. As a result, this approach
is robust to several important categories of missingness patterns, with the most
important being highly angled teats.

\section{Discussion}
\label{sec:discussion}
In this work we have  proposed and evaluated a method for teat length estimation
from udder point clouds with a Gaussian process. In the demonstrations above, we
show that the proposed method improves on the method presented by
\textcite{montes2026}, particularly in the case of udder scans with missing data
and angled teats and is more consistent than our implementation of the methods
described by \textcite{shorten2021}. Additionally, the method we introduce is
automated and computationally efficient making it attractive for phenotyping a
large number of individuals.

Existing methods for estimating teat length rely on gradient-based techniques to
determine the teat base \parencite{shorten2021, Westberg2009, montes2026}. The
method presented by \textcite{shorten2021} fits a regression on the teat radius.
\textcite{Westberg2009} used an algorithm that evaluated the growth of the teat
from the tip to neighboring points. \textcite{chuah2024} used a sliding window
sized to the teat width and determined the distance between the furthest and
closest points in the window to locate and measure the teats. A common
limitation of these methods is that they are sensitive to missing points
throughout the teat segmentation process. Furthermore, these methods typically
define hard-coded thresholds such as defining a specific deviation from a
regression line, a cutoff for teat growth, or window sizes. These fixed
parameters are difficult to generalize across the wide variety of udder
morphologies found in dairy cows. 

In contrast, the udder floor-teat separation in the proposed method relies on
assumptions about surface properties of these two structures rather than the
exact parameter values chosen in the model. Therefore, as long as the assumed
properties hold, the separation will succeed because the udder floor column
space defined to be smooth does not span the teat, which is sharp. This approach
keeps the model flexible to diverse udder morphologies. The teat postprocessing
algorithm presented here is not entirely free of hard-coded constants, with one
example being the use of pixels at least $1/4$ the height of the tip for
determining teat angle (Algorithm~\ref{alg:teat_len}), but this value was not
tuned to optimize outcomes and model performance is not sensitive to it. For
instance, changing this cutoff to 1/8, 1/3 and 1/2 yield similar RMSE as 1/4 on
the test set (7.33, 8.55, and 8.17 mm, respectively).

As mentioned earlier, efficient computation was a primary concern in this work.
The primary mechanism to achieve this was to design the approach so as to only
ever interpolate very smooth functions, which can be done efficiently and
accurately. The methods proposed by \textcite{montes2026} and
\textcite{shorten2021}, in comparison, are computationally expensive because
they require iterating through the point cloud data over discrete intervals and
performing many interpolations along the way.  In the case of
\textcite{montes2026}, the algorithm repeatedly calculates the distance between
pairs of points within areas of increasing radii, dividing this by the increment
in height until the height-distance ratio falls below a specific threshold. That
approach relies on interpolation of the missing points along the teat, which is
slow and can introduce artifacts that lead to inaccurate teat lengths.
Similarly, the method described by \textcite{shorten2021} requires sequentially
calculating the radii of contour levels to fit the regression. Although this
method occasionally yields accurate teat lengths, the failed cases strongly
deviate from the manual annotations because the regression error threshold is
exceeded on the first contours, which may be due to noise near the teat tip or a
regression model fitted with few levels due to missing points.

One limitation of this work is that the model parameters are designed directly
to estimate teat length accurately, rather than to identify the locations of the
teat tip and teat base and other relevant physical markers. Future work could
refine the objective function to locate these landmarks accurately.
Additionally, the dataset consisted of 48 quarters from cows from a single herd.
Therefore, this dataset may not capture the different occlusion patterns or
extreme morphologies. Future research should evaluate the generalizability of
this approach on a larger population of cows. A related question that this work
does not address is uncertainty quantification. While a GP is used to obtain a
reduced-rank basis for the udder function, it is not directly applicable to
assessing the variability of teat length estimators because the most relevant
source of uncertainty is the cutoff at which a gradient in the udder is
sufficiently sharp that it is deemed to be the beginning of the teat (see, for
example, the left front segment of Cow 1081 in Figure
\ref{fig:teat_len_examples}). The design of a method to characterize the
relevant uncertainty in the teat length estimators proposed here is an exciting
avenue for future work.

\printbibliography

\newpage

\appendix

\section{Adaptive cross approximation algorithm} \label{app:aca}

The below algorithm gives a terse and mathematically simplified description of
the ACA that is sufficiently detailed for direct re-implementation. For
mathematical details of the algorithm, we refer readers to
\parencite{bebendorf2003}.

\begin{algorithm}[h!]
\caption{Partially Pivoted Adaptive Cross Approximation}
\label{alg:aca}
\begin{algorithmic}[1]
  \REQUIRE Row and column queries of $M \in \mathbb{R}^{m \times n}$, tolerance $\tau$, max rank $k_{\max}$
  \ENSURE Low-rank factors $U \in \mathbb{R}^{m \times k}$ and $V \in \mathbb{R}^{n \times k}$ such that $M \approx UV^T$
  \STATE $z \gets 0$
  \FOR{$l = 1$ \TO $k_{\max}$}
    \IF{$l = 1$}
      \STATE Initialize $i^*$ and extract row $r \gets M_{i^*, :}$
    \ELSE
      \STATE Find next pivot row $i^*$ from previous column buffer \COMMENT{Get next row index $i^*$}
      \STATE $r \gets M_{i^*, :}$ \COMMENT{Get next row buffer}
      \FOR{$t = 1$ \TO $l-1$}
        \STATE $r \gets r - U_{i^*, t} \, V_{:, t}$ \COMMENT{Remove earlier ranks}
      \ENDFOR
    \ENDIF
    \STATE $j^* \gets \arg\max_j |r_j|$ \COMMENT{Get the next column index $j^*$}
    \STATE $v_{\max} \gets r_{j^*}$
    \IF{$|v_{\max}| < \tau$}
      \RETURN $U_{:, 1:l-1}, V_{:, 1:l-1}$ \COMMENT{Early termination criterion}
    \ENDIF
    \STATE $r \gets r / v_{\max}$ \COMMENT{Normalize row}
    \STATE $c \gets M_{:, j^*}$ \COMMENT{Get the column buffer}
    \FOR{$t = 1$ \TO $l-1$}
      \STATE $c \gets c - V_{j^*, t} \, U_{:, t}$ \COMMENT{Remove earlier ranks}
    \ENDFOR
    \STATE $U_{:, l} \gets c$ \COMMENT{Add the new column to factors}
    \STATE $V_{:, l} \gets r$ \COMMENT{Add the new row to factors}
    \FOR{$t = 1$ \TO $l-1$} 
      \STATE $z \gets z + 2 \left|\langle U_{:, t}, c \rangle\right|
      \left|\langle V_{:, t}, r \rangle\right|$ \COMMENT{Update error tracking quantities}
    \ENDFOR
    \STATE $z \gets z + \|c\|^2 \|r\|^2$
    \IF{$\sqrt{\|c\|^2 \|r\|^2 / z} < \tau$ \AND $l > 1$}
      \RETURN $U_{:, 1:l}, V_{:, 1:l}$ \COMMENT{Convergence criterion}
    \ENDIF
  \ENDFOR
  \RETURN $U, V$
\end{algorithmic}
\end{algorithm}

\newpage

\section{Implementation of \textcite{shorten2021} method} \label{app:shorten}
We implemented the method from \textcite{shorten2021} for teat length estimation
on udder point clouds. This method identifies the teat base at the height above
the teat tip where the observed contour radius exceeds the predicted regression
radius by 3~mm. We used the alpha shape method \parencite{alphashape2021}, with
$\alpha = 2$, to determine the contour levels and the Shoelace formula to
calculate the area within those contours. 

\begin{algorithm}
    \caption{Implementation of \textcite{shorten2021} teat length estimation}
    \label{alg:teat_len_shorten}
    \begin{algorithmic}[1]
        \REQUIRE $\mathcal{P} \subset \mathbb{R}^{3}$ : quarter 3D point cloud ($[x, y, z]$) in mm
        \STATE $t \gets\min_{\mathbf{p} \in \mathcal{P}} (\mathbf{p}_z)$ \COMMENT{Locate the teat tip}
        \STATE $\mathcal{L} = [l_1, l_2, \dots, l_M]^T$ \COMMENT{Front $l_1 =5$ and $l_M = 45$, Rear $l_1 =5$ and $l_M = 20$}
        \STATE $\mathbf{r} = [r_1, r_2, \dots, r_M]^T$
        
        \FOR{$i = 1$ \TO $M$}
            \STATE $h_i \gets t + l_i$
            \STATE $\mathcal{S}_i \gets \{\mathbf{p} \in \mathcal{P} \mid \mathbf{p}_z \le h_i\}$ \COMMENT{Filter points below the height threshold}
            \STATE $A_i \gets \text{area of the 2D contour level of } \mathcal{S}_i$
                \STATE $r_i \gets \sqrt{A_i / \pi}$ \COMMENT{Compute radius of the contour level}
        \ENDFOR
        
        \STATE $\mathbf{X} = [\mathbf{1}, \mathcal{L}]$ 
        \STATE $\bm{\beta} = (\mathbf{X}^T \mathbf{X})^{-1} \mathbf{X}^T \mathbf{r}$
        
        \FOR{$l = 1$ \TO $100$}
            \STATE $\hat{r} \gets l \cdot \bm{\beta}_2 + \bm{\beta}_1$ \COMMENT{Predict the radius from the regression equation}
            \STATE $h \gets t + l$
            \STATE $\mathcal{S} \gets \{\mathbf{p} \in \mathcal{P} \mid \mathbf{p}_z \le h\}$ \COMMENT{Filter points below the height threshold}
            \STATE $A \gets \text{area of the 2D contour level of } \mathcal{S}$ 
            \STATE $r \gets \sqrt{A / \pi}$ \COMMENT{Compute radius of the contour level}
            \IF{$(r- \hat{r}) > 3$} 
                \STATE $h_{\text{teat}} \gets h$
                \STATE \textbf{break}
            \ENDIF
        \ENDFOR
        
        \RETURN $h_{\text{teat}}$
    \end{algorithmic}
\end{algorithm}

\end{document}